\documentclass{emulateapj}

\usepackage{graphicx}
\usepackage{natbib}
\shorttitle{Foreground Removal of 21-cm Observations}
\shortauthors{Cho et al.}

\begin{document}
\title{A Technique for Foreground Subtraction in Redshifted 21 cm Observations} 
\author{Jungyeon Cho\altaffilmark{1}, A. Lazarian\altaffilmark{2}, 
        and Peter T. Timbie\altaffilmark{3}}  

\altaffiltext{1}{Dept. of Astronomy and Space Science,
       Chungnam National University, Daejeon, Korea} 
\altaffiltext{2}{Dept. of Astronomy, University of Wisconsin,
       Madison, WI 53706, USA}
\altaffiltext{3}{Dept. of Physics, University of Wisconsin,
       Madison, WI 53706, USA}

\begin{abstract}
One of the main challenges for future 21 cm observations is to remove foregrounds which are several orders of
magnitude more intense than the HI signal.
We propose a new technique for removing foregrounds of the redshifted 21 cm observations. 
We consider multi-frequency interferometer observations.
We assume that the 21 cm signals in
different frequency channels are uncorrelated and
the foreground signals change slowly as a function of frequency.
When we add the visibilities of all channels, the foreground signals increase roughly by a factor of $\sim N$
because they are highly correlated. However, the 21 cm signals increase by a factor of $\sim \sqrt{N}$ because
the signals in different channels contribute randomly.
This enables us to obtain an accurate shape of the foreground angular power spectrum.
Then, we obtain the 21-cm power spectrum by subtracting the foreground power spectrum obtained this way.
We describe how to obtain the average power spectrum of the 21 cm signal.

\end{abstract}
 
\section{Introduction}
A promising emerging field in cosmology is the proposed use of 21-cm emission from 
neutral hydrogen to trace the evolution of structure in the universe from redshift 0 to 50.  
At redshifts up to about 3, by mapping the three-dimensional intensity field with 10 arc-minute resolution, 
it may be possible to precisely measure the expansion history throughout the transition from 
deceleration to acceleration \citep{chang08, wyithe08,LeobW08,Morales09}.   
For redshifts near 10 several instruments are under construction to search for 21-cm emission from 
the Epoch of Reionization \citep[EoR;][]{Morales04,zaroubisilk05,Backer07}. 
Some authors even suggest that observations at redshifts well above 10 may provide 
very sharp tests of the world model and the composition of the cosmic fluid \citep{LeobZ04}.

Most of the observational effort so far has focused on searching for the EoR signal.  The first goal of current EoR programs is to pin down the redshift of reionization.  Later measurements 
will address questions such as the following:  
What were the first sources of ionizing radiation?  
What is the power spectrum of the HI structure, and how did it evolve?
The first generation of 21-cm EoR observatories may begin to attack
this last question by observing the emission of the neutral hydrogen itself.  The interpretation
of the data in terms of pinning down the UV sources is likely to be much more difficult.

Even at the end of the EoR, there is still enough HI (a few percent of all the hydrogen) in the form of 
self-shielded clumps  to be visible at a brightness temperature contrast of $\Delta T \sim 1$ mK. 
We have some information about these structures since they are seen via UV absorption 
as Damped Lyman Alpha regions, or DLA's \citep{wolfe2005}.
Observation of 21-cm fluctuations from such systems provides 
a tool for understanding structure formation in the post-reionization era.  
These HI regions presumably trace 
the dark matter and hence can be used to study the evolution of the matter power spectrum.
Measurement of the cosmic power spectrum as a function of redshift over the cosmic volume $0.5 < z < 6$ could provide cosmic-variance limited determination of cosmological parameters such as the equation of state of the dark energy and  the neutrino mass \citep{LeobW08}.    

Multiple lines of observational evidence now indicate that dark energy accounts for $\sim$ 70\% of the 
energy density in the universe, but so far we have few
clues as to the physics underlying this phenomenon. 
There are a host of dark energy models and these
can be distinguished by their equations of state. 
A promising approach is to infer the equation of state from the distance-redshift relation in the redshift range of 0 to about 2.   In particular, baryon-acoustic oscillations (BAO) can provide a standard ruler for distance determinations.  The BAO  \lq wiggles' in the matter power spectrum can be measured efficiently using 21-cm intensity mapping \citep{chang08,ansari08,ansari11,seo10} 

To achieve these scientific goals a wide variety of innovative 21-cm telescope designs have been proposed, some prototypes built, 
and some telescopes completed  
  \citep[see][for a review 
of these instruments and their scientific potential]{Morales09}.
However, there are many development tasks needed to advance the state of the art of 21-cm cosmology, 
including calibration strategies, low-noise amplifier development, correlator design, 
examination of array layouts,  radio-quiet site development, and foreground removal.

One of the main difficulties of all 21-cm telescopes is to remove foregrounds which are several orders of magnitude more intense than the HI signal \citep{MoralesB06}. The basic idea is that most foregrounds have a smooth power-law frequency dependence in contrast with the HI signal. The exception is Galactic radio recombination lines which occur at known frequencies and can be excised. The main
foregrounds are the Galactic synchrotron radiation and extragalactic point sources. Other sources like
free-free electron emission ({\sl i.e.}~Bremsstrahlung) are much less intense and have power-law spectra
similar to the first two components. 

Other foreground removal techniques have been proposed as well.  Morales {\it et al.}  (\citeyear{MoralesB06}) 
review these techniques and show how to correct the errors they cause in the recovered power spectrum.  
Liu {\it et al.} (\citeyear{liu09}) show that performing foreground removal in Fourier space eliminates errors 
that arise from frequency-dependent beam patterns (mode mixing).  
Liu {\it et al.} (\citeyear{liu11}) have developed a technique for estimating the power spectrum 
in the presence of foregrounds without introducing noise or bias to the power spectrum.

Simulations by Bowman, 
Morales and Hewitt (\citeyear{BowmanM09})
for the Murchison Widefield Array 
remove the contribution from bright identified sources and then in each RA-dec
pixel fit and subtract a polynomial in $\nu$.  This two-step strategy allows for the observation of
the HI signal at the time of reionization $(z\approx 8)$. 
Similar foreground removal simulations have been applied to 
candidate dark energy observatories;   Ansari {\it et al.} \citeyear{ansari11} show how foregrounds 
might be separated from the HI signal based  on their differing spectral signatures as measured 
by an interferometric array of dishes or cylinder telescopes.  

Below we discuss a different idea for removing foregrounds which is based on using the
power spectrum. The physical basis of the idea is that at different frequencies the
fluctuations of redshifted HI are not correlated, as we sample completely different
regions of  HI, while the foreground fluctuations are correlated as they arise from
the same galactic inhomogeneities, e.g. turbulent fluctuations as discussed in Cho \& Lazarian \citeyear{CL02,CL10}. 
However, unlike Cho \& Lazarian (2010) the suggested analysis does
not require assumptions about the power spectrum of fluctuations.

In \S2 we discuss the technique that we propose, 
in \S3 the applicability of our technique, and in \S4 we summarize
our results.

\begin{figure}
\center
\includegraphics[angle=0,width=0.49\textwidth]{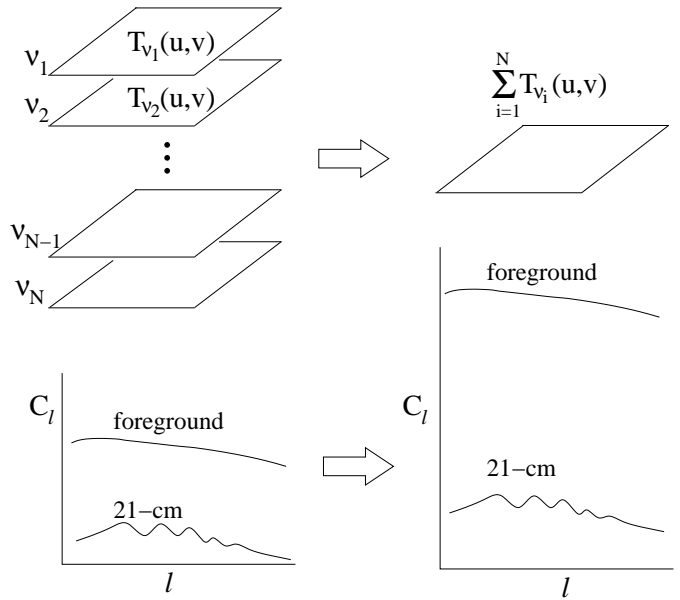}
\caption{Method. {\it Left:} Observation are done in $N$ different frequency channels ($\nu=\nu_1, ..., \nu_N$). 
      The foreground signals
      are several orders of magnitude stronger than the 21-cm signal (low-left panel).
      We assume that the 21-cm signals from different channels are uncorrelated.
      However, we assume the foreground signals from different channels are highly correlated.
      {\it Right:} When we add all $T_{\nu_i}(u,v)$'s, the foreground signals are enhanced by a factor of
      $\sim N$, while the 21-cm signal is enhanced by a factor of $\sim \sqrt{N}$.
      In terms of angular spectrum, the foreground spectrum goes up by a factor of $\sim N^2$ and
      the 21-cm spectrum goes up by $\sim N$.
      The angular spectrum of $T_{avg}(u,v)\equiv \sum_i T_{\nu_i}(u,v)/N$ will be 
      $\sim C_l^{for} + C_l^{21cm}/N$, where
      `for' and `21-cm' stand for the foreground and the redshifted 21-cm, respectively.
       Therefore, the spectrum of $T_{avg}(u,v)$, $C_l^{avg}$, should represent 
       an accurate shape of the foreground spectrum. 
       We can obtain the 21-cm spectrum by subtracting the foreground spectrum. }
      \label{fig1}
\end{figure}

\section{Method}

\subsection{Assumptions}
We assume the following:
\begin{enumerate}
\item  The observations are made at $N$ different frequency channels ($\nu_i$, $i$=$1$,..., $N$)
       and data are collected in $uv$ coordinates.
       We assume that $\delta \nu $ ($\equiv \nu_{i+1}-\nu_i$) is constant. 
       We denote the visibility at channel $i$ as $ T_{\nu_i}(u,v)$.
       
\item Observations are made for a small patch on the sky. We use $(\theta_x, \theta_y)$ to denote
      a position on the sky plane. We use either $(k_x, k_y)$ or $(u,v)$ to denote
      a position in Fourier space. Note that 
      $l = 2\pi (u^2+v^2)^{1/2}$ \citep{white99}
      and $C_l^{\nu_i}\propto  |T_{\nu_i}(u,v)|^2
      = (1/N_{sum})\sum_{ l/2\pi}  |T_{\nu_i}(u,v)|^2$,
      where the summation is done for $ (l/2\pi) \leq \sqrt{u^2+v^2} <  (l/2\pi)+(\Delta l/2\pi)$ and
       $N_{sum}$ is the number of modes used for the summation. 
      
\item The 21-cm signals from different frequency channels are uncorrelated.

\item The cross-correlation between the foreground and the 21-cm signals is negligible.
      This is not a trivial issue if the 21-cm signal is much weaker than the foreground one.
      We will come back to this in \S\ref{sect:diss}. 
      
\item The noise level is lower than the 21-cm signal. See also discussions in \S\ref{sect:diss}.

\end{enumerate}

\subsection{Basic idea}

We can obtain an accurate shape of the foreground angular power spectrum in the following way.
 Suppose that we have multi-channel observation data in $uv$ coordinates (left panel of Fig. 1).
We stack all the $uv$ plane data into a single file (right panel of Fig. 1). 
That is, we calculate $\sum_{i=1}^{N} T_{\nu_i}(u,v)$.
(If the observations are done in real space, we first obtain $\sum_{i=1}^{N} T_{\nu_i}(\theta_x, \theta_y)$ 
      and then we perform a Fourier transform.)
In the stacked map, since the foreground signals in different channels are correlated, they are enhanced by a factor of
$\sim N$. However, the noise and the 21-cm signals are enhanced by a factor of $\sqrt{N}$ because they contribute randomly.
Therefore, 
the contrast between the foreground and the noise/21-cm angular spectra
will be increased by $\sim N$ times. 
In this way we can obtain an accurate shape of the foreground power spectrum.
We define $T_{avg}(u,v)$ as
     \begin{equation}
        T_{avg}(u,v)\equiv \frac{1}{N} \sum_{i=1}^{N} T_{\nu_i}(u,v) .  \label{eq:Tavg}
     \end{equation}
     
What, then,  does $T_{avg}(u,v)$ mean?
If $T_{\nu}(u,v)$ varies slowly as a function of $\nu$, we can show that
the obtained foreground power spectrum is a good representation for the power spectrum at the central channel
($\nu = \nu_{(N+1)/2}$).
Suppose that N is an {\it odd} number. Then we can write
    \begin{equation}
       T_{\nu_i}^{for}(u,v)= T_{\nu_{(N+1)/2}}^{for}(u,v)+ a \Delta \nu,   \label{eq:linear}
   \end{equation}
   where $a$  is a constant, $\Delta \nu = \nu_i-\nu_{(N+1)/2}$ and the superscript `for' denotes the foreground.
   We assume $|a \Delta \nu | \ll |T_{\nu_i}^{for} |$ and we ignore the second-order quantities.
   {}From Eqs.~(\ref{eq:Tavg}) and (\ref{eq:linear}), we have 
   $ 
      \frac{1}{N}\sum_{i=1}^{N} T_{\nu_i}(u,v) \approx T_{\nu_{(N+1)/2}}^{for}(u,v),
   $ 
   which results in
   \begin{equation}
       C_l^{avg} \approx C_l^{\nu_{(N+1)/2}, for},
   \end{equation}
   where $C_l^{avg}$ and $C_l^{\nu_i, for}$ are the angular power spectra of $T_{avg}(u,v)$
   and $T_{\nu_i}^{for}(u,v)$, respectively.

The foreground power spectrum ($\approx C_l^{avg}$), as well as the foreground map ($\approx T_{avg}(u,v)$), 
can be used for other observations.
For example, when a future low-noise single channel observation becomes
available at $\nu=\nu_{(N+1)/2}$, we can use $C_l^{avg}$ or $T_{avg}(u,v)$ to subtract the foreground.
It is also possible to use them for future high angular resolution low-noise observations.
Cho \& Lazarian \citeyearpar{CL10} argued that when accurate foreground observations
are available for a certain range of $l$'s, we can extrapolate the foreground
spectrum to a range of $l$ where no 
foreground information is available. Therefore, if a very high angular resolution
single-frequency observation is performed in the future, we can extrapolate the foreground spectrum (i.e.~$C_l^{avg}$) to
higher values of $l$ and extract the 21-cm information on the small scales.

\subsection{Proposed techniques}

{\bf Case 1: same power spectrum shape for all channels}---In this case, we 
   can easily obtain $C_l^{\nu_i, 21cm}$. Note that, at this moment, we know only the shape (the $l$-dependence)
   of $C_l^{\nu_i, for}$. The amplitude can be different in each frequency channel and we do not know it yet.
   We can determine the amplitude by fitting or by the following
   method.
   Suppose that the value of
   \begin{equation}
         \sum_{l} | \log(C_l^{\nu_i}) - \log( \alpha C_l^{avg})|^2
   \end{equation} 
   has a minimum value at $\alpha=\alpha_i$.
   Then we have
   \begin{equation}
     C_l^{\nu_i, 21cm} \approx C_l^{\nu_i} - \alpha_i C_l^{avg}.
   \end{equation}

{\bf Case 2: linear dependence of T(u,v) on $\nu$}---In this case, as we discussed
in the previous subsection, we have $C_l^{avg} =C_l^{\nu_{(N+1)/2, for}}$.
The 21-cm spectrum at $\nu=\nu_i$ is
\begin{equation}
   C_l^{\nu_i, 21cm}=C_l^{\nu_i}-C_l^{\nu_{i, for}}.
\end{equation}
{}From Eq.~(\ref{eq:linear}), we have
   \begin{equation}
       C_l^{\nu_i, for}=C_l^{\nu_{(N+1)/2}, for} + A \Delta \nu,
   \end{equation}
   where $A=
   Re[ 2\sum_{ l/2\pi} T_{\nu_{(N+1)/2}}^{for}(u,v) a^*]$.
   Here `*' denotes the complex conjugate and $Re[...]$ stands for the real part.
   Therefore,
the average spectrum of the 21-cm signal over $N$ channels is
   \begin{equation}
      \frac{1}{N}\sum_{i=1}^{N} C_l^{\nu_i,21cm} =\frac{1}{N} \left( \sum_{i=1}^{N} C_l^{\nu_i} \right) - C_l^{avg}. 
      \label{eq:caseB}
   \end{equation}

{\bf Case 3: quadratic dependence of T(u,v) on $\nu$}-In this more general case,
we can obtain the average 21-cm spectrum. Suppose that $N$={\it even} for simplicity.
Then we can write
\begin{equation}
   T_{\nu_i}^{for}(u,v)=f_0+ a \Delta \nu + b \Delta \nu^2,   \label{eq:quad}
\end{equation}
where $f_0$ is a visibility at $\nu_0=(\nu_{N/2}+\nu_{1+N/2})/2$ (not observed), 
$a$ and $b$ are constants, 
$\Delta \nu = \nu_i-\nu_0 =[i-(N+1)/2] \delta \nu$, and $\delta \nu=\nu_{i+1}-\nu_i$.
We note that
\begin{eqnarray}
    | T_{avg}(u,v)|^2  
    =\left| f_0 + \frac{b}{2N} \Delta \nu_0^2 \sum_{j=1}^{N/2} (2j-1)^2 \right|^2, \nonumber \\
    =| f_0 |^2 + \frac{1}{N} \Delta \nu_0^2 \sum_{j=1}^{N/2} (2j-1)^2
                                             Re [ f_0 b^* ].  \label{eq:avg}
\end{eqnarray}
On the other hand, we have
\begin{eqnarray}
  \sum_{i=1}^{N/2} |T_{\nu_i}(u,v)+T_{\nu_{N-i+1}}(u,v)|^2 \mbox{~~~~~~~~}  \label{eq:pairs} \\  
  =\sum_{j=1}^{N/2} \left| 2f_0 + \frac{b}{2} (2j-1)^2 \delta \nu^2 \right|^2    
     +\sum_{i=1}^{N/2} | T_{\nu_i}^{21cm} + T_{\nu_{N-i+1}}^{21cm} |^2 \nonumber \\
  = 2N | f_0 |^2 + 2 \delta \nu^2 \sum_{j=1}^{N/2} (2j+1)^2 
                                             Re [ f_0 b^*] 
            + \sum_{i=1}^{N}|T_{\nu_i}^{21cm}|^2    \nonumber
\end{eqnarray}
where we dropped the cross-correlation terms between the 21-cm and the foreground signals (see \S\ref{sect:diss}).
If we subtract Eq. (\ref{eq:avg}) times $2N$ from Eq. (\ref{eq:pairs}), we have
  $\sum_{i=1}^{N}|T_{\nu_i}^{21cm}|^2$.
Therefore, the average spectrum of the 21-cm signal is
\begin{equation}
   \frac{1}{N}\sum_{i=1}^{N} C_l^{\nu_i,21cm}=\frac{1}{N}(\sum_{i=1}^{N/2} D_l^{\nu_i}) - 2C_l^{avg},
   \label{eq:caseC}
\end{equation}
where $D_l^{\nu_i}$ is the angular spectrum of $T_{\nu_i}(u,v)+T_{\nu_{N-i+1}}(u,v)$ 
\footnote{More precisely, the right hand side is multiplied by $N/(N-2)$. 
          Similarly, the right hand side of Eq.~(\ref{eq:caseB}) is multiplied by $N/(N-1)$.}.

\section{Discussion}   \label{sect:diss}

Our technique critically depends on the assumption of slow variation of the $T_{\nu_i}^{for}(u,v)$
over frequency channels.
This assumption may need further investigation.
However, we believe that this is a good assumption due to the following reasons.
First of all, we know that the major foreground, which is the synchrotron foreground, is caused by MHD turbulence
\citep[see the statistical description of fluctuations in][]{LazP12}. 
The variations
with frequency from an elementary synchrotron emitting volume are given by
  $  
   \propto n_{0} B_{\perp}^{2} \nu^{(1-\alpha)/2}
  $  
where $B_{\perp}$ is the magnetic field strength perpendicular to
the line of sight and the number density of cosmic rays has a power law distribution 
$n(E)dE=n_{0}E^{-\alpha}dE \sim n_0 E^{-3}dE$. 
Therefore we expect the spectral components to change 
as $\propto (\nu/\nu_0)^{-1}$. 
This dependence can be compensated over the frequency interval of $\Delta \nu=(N-1)\delta \nu$
and the residual emission, e.g. arising from other subdominant components of the foreground, 
may be treated with the suggested technique. The dependence of the other foreground
components, e.g. free-free, dust and spinning dust \citep[see][for a review]{LazF03} on the frequency
is weak and therefore the linear/quadratic expansion in frequencies in Eq. (\ref{eq:linear})/(\ref{eq:quad}) should be also
valid. 
If we observe, for example, 610 MHz, the 21-cm signal is believed to be decorrelated 
if $\Delta \nu > 0.5$ MHz (see for example Ghosh et al. \citeyear{Ghosh11}). 
If we use 12 channels, we have $\Delta \nu/\nu_0 =5.5/610 \approx 0.01$.
Therefore, any third or higher order effects should be very small. 
For synchrotron, since $ (\nu/\nu_0)^{-1}=\sum_{n=0}^{\infty} (1-\nu/\nu_0)^n$,
the ratio of the third to the second term of the expansion is $\sim O(\nu/\nu_0-1) \sim 0.01$.
For free-free
the frequency dependence is weaker and truncation error will be smaller. 

In this paper we assume that the cross-correlation between the foreground and the 21-cm signals is negligible.
This needs further verification.
 In general, we can write
      $C_l=C_l^{for}+C_l^{21cm}+C_l^{cc}$, where 
      \begin{equation}
         C_l^{cc}=Re \left[ \frac{2}{N_{sum}} \sum_{ l/2\pi} T^{for}(u,v)T^{21cm}(u,v)^* \right] \label{eq:cc}
      \end{equation}
      is the cross-correlation term.
      When the 21-cm signal is $\gamma$ times smaller than the foreground one 
      (i.e.~$T^{21cm} \sim T^{for}/\gamma$), the cross-correlation term $C_l^{cc}$ can be larger than
      $C_l^{21cm}$ if the number of modes used for the summation in Eq.~(\ref{eq:cc}) is less than
      $\sim \gamma^2$. 
      
      However, if we use Eq.~(\ref{eq:caseC}) or (\ref{eq:caseB}), we may circumvent
      this difficulty. Let us consider the more general case of Eq.~(\ref{eq:caseC}).
      The cross-correlation term that will appear additionally in the right-hand side of Eq.~(\ref{eq:caseC}) is
      \begin{eqnarray}
         C_l^{cc} = 
         Re \left[ \frac{2}{N_{sum}} \sum_{ l/2\pi} 
          ( \mathcal{A}-\mathcal{B} ) \right],  \label{eq:cc2} \\
          \mathcal{A}=\frac{1}{N}\sum_{i=1}^{N/2} \left[ 
          ( 2 f_0+2b\Delta \nu^2) ( T_{\nu_i}^{21cm} +T_{\nu_{N-i+1}}^{21cm} )^*  \right]  ,\\
           \mathcal{B}= \frac{2}{N^2} \left( \sum_{i=1}^{N} ( f_0+b \Delta \nu^2 ) \right)   
                \left(  \sum_{i=1}^{N} T_{\nu_i}^{21cm*} \right) , \\
        | \mathcal{A-B} | \lesssim 2|b| \Delta \nu_{max}^2 |T_{\nu_i}^{21cm}| /\sqrt{N},
      \end{eqnarray}
      where the variables are defined similarly (i.e.~$\Delta \nu = \nu_i-\nu_0$, N=even,...),
       `$\mathcal{A}$'  and `$\mathcal{B}$' are from the first and the second terms 
       on the right-hand side of Eq.~(\ref{eq:caseC}), respectively,
      $\Delta \nu_{max}=\nu_N-\nu_0$, and we dropped `(u,v)' for simplicity.
      Note that the terms containing $f_0$ cancel out.
      Therefore, if 
      \begin{equation}
         2b\Delta \nu_{max}^2/\sqrt{NN_{sum}} < |T_{\nu_i}^{21cm}|,  \label{eq:cond_cc}
      \end{equation}
      we can neglect the cross-correlation term.
      If the cross-correlation term is indeed larger than the 21-cm spectrum $C_l^{21cm}$ for any reason,
      our technique %
      can reveal the cross-correlation spectrum.
      Even in this case, it may not be difficult to infer the shape of $C_l^{21cm}$ from the shape of $C_l^{cc}$.

Our technique is different from standard polynomial-fitting approaches \citep[e.g.,][]{McQuinn06,liu09}.
First, although we assume a polynomial dependence of the foregrounds on frequency, 
we do not need the actual fitting stage. 
Therefore our technique can be used as a synergistic technique to previous standard suggestions.
Second, since $C_l \propto \sum |T^2|$, we expect that the high-order frequency dependence of $T$ may cancel out
during the summation.
Therefore, 
in the power spectrum we may have a smoother signal which is much easier to constrain.
Third, our technique can give a useful insight for removing noise.
Suppose that noise signals are larger than or comparable to the 21-cm signals and that 
the 21-cm signals at $\nu$ and $\nu^{\prime}$ are correlated if 
$|\nu -\nu^\prime| \lesssim \Delta \nu = \nu_{i+1}-\nu_{i}$.
In this case 
our technique will give an average spectrum of the noise and the 21-cm signals, $\sim C_l^{noise}+C_l^{21cm}$.
Now, we add $M-1$ new channels between each $\nu_i$ and $\nu_{i+1}$ ($i=1,...,N$), so that there are 
a total of $NM$ channels. Here we assume that the total bandwidth stays roughly the same and
the bandwidth of each channel is now $\sim M$ times narrower than before.
Then we apply our technique again. 
Note that the 21-cm signals are correlated in adjacent channels, while the noise signals are not.
The newly obtained average spectrum is now proportional to $\sim  C_l^{noise}+C_l^{21cm}/\sqrt{M}$.
Therefore, by comparing the two results, we may extract
the spectrum of the 21-cm signals. 
For example, if the noise and the 21-cm signals have different dependence on $l$ (or $k$),
we can apply the technique in Cho \& Lazarian (2010).

\section{Summary}
We have described a new technique to remove the foreground from
redshifted 21-cm observations.
We have assumed that the foreground signals change slowly as the frequency changes: we have assumed
up to a quadratic dependence of $T_{\nu_i}^{for}(u,v)$ on frequency.
We obtain the foreground spectrum by simply adding all the observed $T_{\nu_i}(u,v)$.
If the observations are done in real space, we add all the $T_{\nu_i}(\theta_x,\theta_y)$
in real space first and then we  perform the Fourier transform.
When we add $T_{\nu_i}(u,v)$ of all channels, the foreground spectrum goes up by a factor of $\sim N^2$
because they are highly correlated. However, the 21-cm spectrum goes up by a factor of $\sim N$ because
the signals in different channels contribute randomly.
This way, we can obtain an accurate shape of the foreground power spectrum.
Then, we obtain the 21-cm power spectrum by subtracting the foreground power spectrum obtained this way.
We have described how to obtain the average 21-cm spectrum (Eqs.~[\ref{eq:caseB}] and [\ref{eq:caseC}]).
We have derived the condition for neglecting the cross-correlation term (Eq.~[\ref{eq:cond_cc}]).

\acknowledgements
J.C.'s work was financially supported by the National Research Foundation of Korea (NRF)
(2011-0012081).
AL acknowledges the support of
the NSF Center for Magnetic Self-Organization, the NASA grant NNX11AD32G, and the NSF grant AST 0808118. 
PT is partly supported by NSF grant AAG 	
0908900.


\begin{thebibliography}{23}
\expandafter\ifx\csname natexlab\endcsname\relax\def\natexlab#1{#1}\fi

\bibitem[{{Ansari} {et~al.}(2008){Ansari}, {Le Goff}, {Magneville}, {Moniez},
  {Palanque-Delabrouille}, {Rich}, {Ruhlmann-Kleider}, \&
  {Y{\`e}che}}]{ansari08}
{Ansari}, R., {Le Goff}, J., {Magneville}, C., {Moniez}, M.,
  {Palanque-Delabrouille}, N., {Rich}, J., {Ruhlmann-Kleider}, V., \&
  {Y{\`e}che}, C. 2008, ArXiv e-prints (0807.3614)

\bibitem[{{Ansari} {et~al.}(2011){Ansari}, {Campagne}, {Colom}, {Le Goff},
  {Magneville}, {Martin}, {Moniez}, {Rich}, \& {Y{\`e}che}}]{ansari11}
{Ansari}, R., {et~al.} 2011, ArXiv e-prints (1108.1474)

\bibitem[{{Backer} {et~al.}(2007){Backer}, {Parsons}, {Bradley}, {Parashare},
  {Gugliucci}, {Mastrantonio}, {Herne}, {Lynch}, {Wright}, {Werhimer},
  {Carilli}, {Datta}, \& {Aguirre}}]{Backer07}
{Backer}, D.~C., {et~al.} 2007, in Bulletin of the American Astronomical
  Society, Vol.~38, Bulletin of the American Astronomical Society, 967

\bibitem[{{Bowman} {et~al.}(2009){Bowman}, {Morales}, \& {Hewitt}}]{BowmanM09}
{Bowman}, J.~D., {Morales}, M.~F., \& {Hewitt}, J.~N. 2009, \apj, 695, 183

\bibitem[{Chang {et~al.}(2008)Chang, Pen, Peterson, \& McDonald}]{chang08}
Chang, T.-C., Pen, U.-L., Peterson, J.~B., \& McDonald, P. 2008, Phys. Rev.
  Lett., 100, 091303

\bibitem[{{Cho} \& {Lazarian}(2002)}]{CL02}
{Cho}, J., \& {Lazarian}, A. 2002, \apjl, 575, L63

\bibitem[{{Cho} \& {Lazarian}(2010)}]{CL10}
---. 2010, \apj, 720, 1181

\bibitem[{{Ghosh} {et~al.}(2011){Ghosh}, {Bharadwaj}, {Ali}, \&
  {Chengalur}}]{Ghosh11}
{Ghosh}, A., {Bharadwaj}, S., {Ali}, S.~S., \& {Chengalur}, J.~N. 2011, \mnras,
  411, 2426

\bibitem[{{Lazarian} \& {Finkbeiner}(2003)}]{LazF03}
{Lazarian}, A., \& {Finkbeiner}, D. 2003, New Astron. Rev., 47, 1107

\bibitem[{{Lazarian} \& {Pogosyan}(2012)}]{LazP12}
{Lazarian}, A., \& {Pogosyan}, D. 2012, ApJ, 745, 5

\bibitem[{{Liu} \& {Tegmark}(2011)}]{liu11}
{Liu}, A., \& {Tegmark}, M. 2011, \prd, 83, 103006

\bibitem[{{Liu} {et~al.}(2009){Liu}, {Tegmark}, {Bowman}, {Hewitt}, \&
  {Zaldarriaga}}]{liu09}
{Liu}, A., {Tegmark}, M., {Bowman}, J., {Hewitt}, J., \& {Zaldarriaga}, M.
  2009, \mnras, 398, 401

\bibitem[{{Loeb} \& {Wyithe}(2008)}]{LeobW08}
{Loeb}, A., \& {Wyithe}, J.~S.~B. 2008, Physical Review Letters, 100, 161301

\bibitem[{{Loeb} \& {Zaldarriaga}(2004)}]{LeobZ04}
{Loeb}, A., \& {Zaldarriaga}, M. 2004, Physical Review Letters, 92, 211301

\bibitem[{{McQuinn} {et~al.}(2006){McQuinn}, {Zahn}, {Zaldarriaga},
  {Hernquist}, \& {Furlanetto}}]{McQuinn06}
{McQuinn}, M., {Zahn}, O., {Zaldarriaga}, M., {Hernquist}, L., \& {Furlanetto},
  S.~R. 2006, \apj, 653, 815

\bibitem[{{Morales} {et~al.}(2006){Morales}, {Bowman}, \&
  {Hewitt}}]{MoralesB06}
{Morales}, M.~F., {Bowman}, J.~D., \& {Hewitt}, J.~N. 2006, \apj, 648, 767

\bibitem[{{Morales} \& {Hewitt}(2004)}]{Morales04}
{Morales}, M.~F., \& {Hewitt}, J. 2004, \apj, 615, 7

\bibitem[{{Morales} \& {Wyithe}(2010)}]{Morales09}
{Morales}, M.~F., \& {Wyithe}, J.~S.~B. 2010, \araa, 48, 127

\bibitem[{{Seo} {et~al.}(2010){Seo}, {Dodelson}, {Marriner}, {Mcginnis},
  {Stebbins}, {Stoughton}, \& {Vallinotto}}]{seo10}
{Seo}, H.-J., {Dodelson}, S., {Marriner}, J., {Mcginnis}, D., {Stebbins}, A.,
  {Stoughton}, C., \& {Vallinotto}, A. 2010, \apj, 721, 164

\bibitem[{{White} {et~al.}(1999){White}, {Carlstrom}, {Dragovan}, \&
  {Holzapfel}}]{white99}
{White}, M., {Carlstrom}, J.~E., {Dragovan}, M., \& {Holzapfel}, W.~L. 1999,
  \apj, 514, 12

\bibitem[{Wolfe {et~al.}(2005)Wolfe, Gawiser, \& Prochaska}]{wolfe2005}
Wolfe, A., Gawiser, E., \& Prochaska, J.~X. 2005, Annu. Rev. Astron.
  Astrophys., 43, 861

\bibitem[{Wyithe {et~al.}(2008)Wyithe, Loeb, \& Geil}]{wyithe08}
Wyithe, S., Loeb, A., \& Geil, P. 2008, \mnras, 383, 1195

\bibitem[{{Zaroubi} \& {Silk}(2005)}]{zaroubisilk05}
{Zaroubi}, S., \& {Silk}, J. 2005, \mnras, 360, L64

\end{thebibliography}

\end{document}